\documentclass[aps,pra,showpacs,twocolumn,superscriptaddress]{revtex4-1}

\usepackage{hyperref}
\usepackage{graphicx}
\usepackage{amsmath,amssymb,latexsym,amsfonts,subfigure,bbm}
\usepackage{dsfont,relsize,color,microtype}
\newcommand{\cc}[1]{\textcolor{black}{#1}}

\begin{document}
\setlength{\tabcolsep}{1pt}
\title{Breaking of $\mathcal{PT}$--symmetry in bounded and unbounded scattering systems}

\author{Philipp Ambichl}
\email{philipp.ambichl@tuwien.ac.at}
\affiliation{Institute for Theoretical Physics, Vienna University of Technology, A--1040 Vienna, Austria, EU}

\author{Konstantinos G.~Makris}
\affiliation{Institute for Theoretical Physics, Vienna University of Technology, A--1040 Vienna, Austria, EU}
\affiliation{Department of Electrical Engineering, Princeton University, Princeton, New Jersey 08544, USA}

\author{Li Ge}
\affiliation{\textls[-18]{Department of Engineering Science and Physics, College of Staten Island, CUNY, New York 10314, USA}}

\author{Yidong Chong}
\affiliation{Division of Physics and Applied Physics, Nanyang Technological University, Singapore 637371, Singapore}

\author{A. Douglas Stone}
\affiliation{Department of Applied Physics, Yale University, New Haven, Connecticut 06520, USA}

\author{Stefan Rotter}
\email{stefan.rotter@tuwien.ac.at}
\affiliation{Institute for Theoretical Physics, Vienna University of Technology, A--1040 Vienna, Austria, EU}

\begin{abstract}
$\mathcal{PT}$--symmetric scattering systems with balanced gain and loss can undergo a symmetry-breaking transition in which the eigenvalues of the non-unitary scattering matrix change their phase shifts from real to complex values. We relate the $\mathcal{PT}$--symmetry breaking points of such an unbounded scattering system to those of underlying bounded systems. In particular, we show how the $\mathcal{PT}$--thresholds in the scattering matrix of the unbounded system translate into analogous transitions in the Robin boundary conditions of the corresponding bounded systems. Based on this relation, we argue and then confirm that the $\mathcal{PT}$--transitions in the scattering matrix are, under very general conditions, entirely insensitive to a variable coupling strength between the bounded region and the unbounded asymptotic region, a result which can be tested experimentally and visualized using the concept of Smith charts.
\end{abstract}

\date{\today}


\maketitle

One of the postulates of quantum mechanics is that Hamiltonians are Hermitian operators, which is justified by the need for real eigenvalue spectra.  Some years ago, Bender and co-authors \cite{BenBoe1998,BenBroJon2002} pointed out that a whole class of ``$\mathcal{PT}$--symmetric'' Hamiltonians also possess real spectra, despite being non-Hermitian \cite{Mos2002,LevZno2000}. This insight has initiated a remarkable surge of research activity, much of it focusing on the ability of such Hamiltonians to undergo ``spontaneous $\mathcal{PT}$--symmetry breaking'' transitions between real and complex (conjugate-pair) eigenvalues \cite{BenBroJon2002,BenBroJon2007,BenDunMei1999,Jon1999,KlaGueMoi2008,Hei2012}.  Although the applicability of $\mathcal{PT}$--symmetric Hamiltonians to quantum mechanics remains speculative, it was realized beginning in 2007 that such Hamiltonians can be studied using optical waveguides and waveguide lattices, where $\mathcal{PT}$--symmetric non-Hermiticity may be implemented with spatially balanced gain and loss \cite{GanMakChr2007,MakGanChr2008,MakGanChr2010}.  Based on these ideas, the $\mathcal{PT}$--breaking transition has now been demonstrated in a variety of experimental systems \cite{RütMakGan2010,GuoSalDuc2009,BitDieGün2012,SchLiZhe2011,RegBerMir2012,BenBerPar2013,peng2013} and was predicted to occur in several more \cite{ChoLiSto2011,LieGeCer2012,ramezani,graefe,prosen,tsironis,kreibich,jensen}.

The use of optical waveguides was motivated by the fact that the Schr\"odinger equation formally maps onto the paraxial equation of diffraction describing propagation of transverse waveguide mode envelopes \cite{GanMakChr2007,MakGanChr2008}, with the $z$-axis playing the role of the time variable, and the $\mathcal{PT}$--breaking transition corresponding to that of the one or two-dimensional {\it bounded} Schr\"odinger problem in the transverse direction.  Subsequently, several authors studied electromagnetic scattering for the Helmholtz equation in an {\it unbounded} domain with a $\mathcal{PT}$--symmetric scatterer, having balanced gain and loss around at least one symmetry plane. For these systems which, in general, do not map onto a bounded Schr\"odinger problem \cite{Mos2009,Mos2011,CanDedVen2007,Lon2010,ChoLiSto2011,Scho2010,LinRamEic2011}, it was recently pointed out \cite{ChoLiSto2011,GeChoSto2012} that the relevant $\mathcal{PT}$--transition occurs in the  eigenvalues of the scattering ($S$) matrix, which relates incoming to outgoing flux channels. Specifically, it was shown that when the $S$ matrix undergoes $\mathcal{PT}$--symmetry breaking, its eigenvalues go from unimodular values, lying on the complex unit circle, to inverse-conjugate pairs of complex values.
This $\mathcal{PT}$--transition in the scattering problem thus raises the question of what relation it might have to a corresponding transition in bounded Hamiltonian systems of the type studied previously.  A recent study of conservation laws in such $\cal{PT}$--scattering systems suggests that a connection to the finite system with Dirichlet boundary conditions exists \cite{GeChoSto2012}.

In this article, we derive an explicit relationship between the $\mathcal{PT}$--breaking transitions of bounded and unbounded systems. Specifically, we show that the regions of the phase diagram in which an unbounded system's $S$ matrix has either unimodular eigenvalues ($\mathcal{PT}$--unbroken phase) or eigenvalues with a modulus different from one ($\mathcal{PT}$--broken phase) correspond exactly to the regions in which a whole \textit{family} of associated bounded systems possess specific $\mathcal{PT}$--symmetric or non-$\mathcal{PT}$--symmetric Robin-type boundary conditions (BC), respectively. This entry will also provide an interesting way to introduce Smith 
charts, familiar to microwave practitioners, into our description.
The importance of Robin BC for relating the bounded and unbounded problems was noted earlier by Smilansky \textit{et al.}~\cite{smilansky1,smilansky2}, Robin BC in the context of $\mathcal{PT}$--symmetry were studied by Krej\v{c}i\v{r}\'{\i}k \textit{et al.}~\cite{KreBilZno2006,Kre2008,KreSie2010,HerKreSie2011};
the correspondence which we demonstrate in the following has, however, not been shown previously.
Although our proof will be presented in the context of the electromagnetic Helmholtz equation, it is also applicable to the Schr\"odinger equation and to other linear wave equations. We also find the surprising, but closely related result that the transition line at which the $S$ matrix undergoes $\mathcal{PT}$--breaking is unchanged by adding arbitrary $\mathcal{P}$--symmetric layers (``mirrors") outside the original scatterer. In other words, the transition is determined entirely by the $\mathcal{PT}$--symmetric inner portion of the optical structure.

\begin{figure}[!b]
  \begin{center}
    \hspace*{-1.25cm}
    \includegraphics[angle=0, scale=0.145]{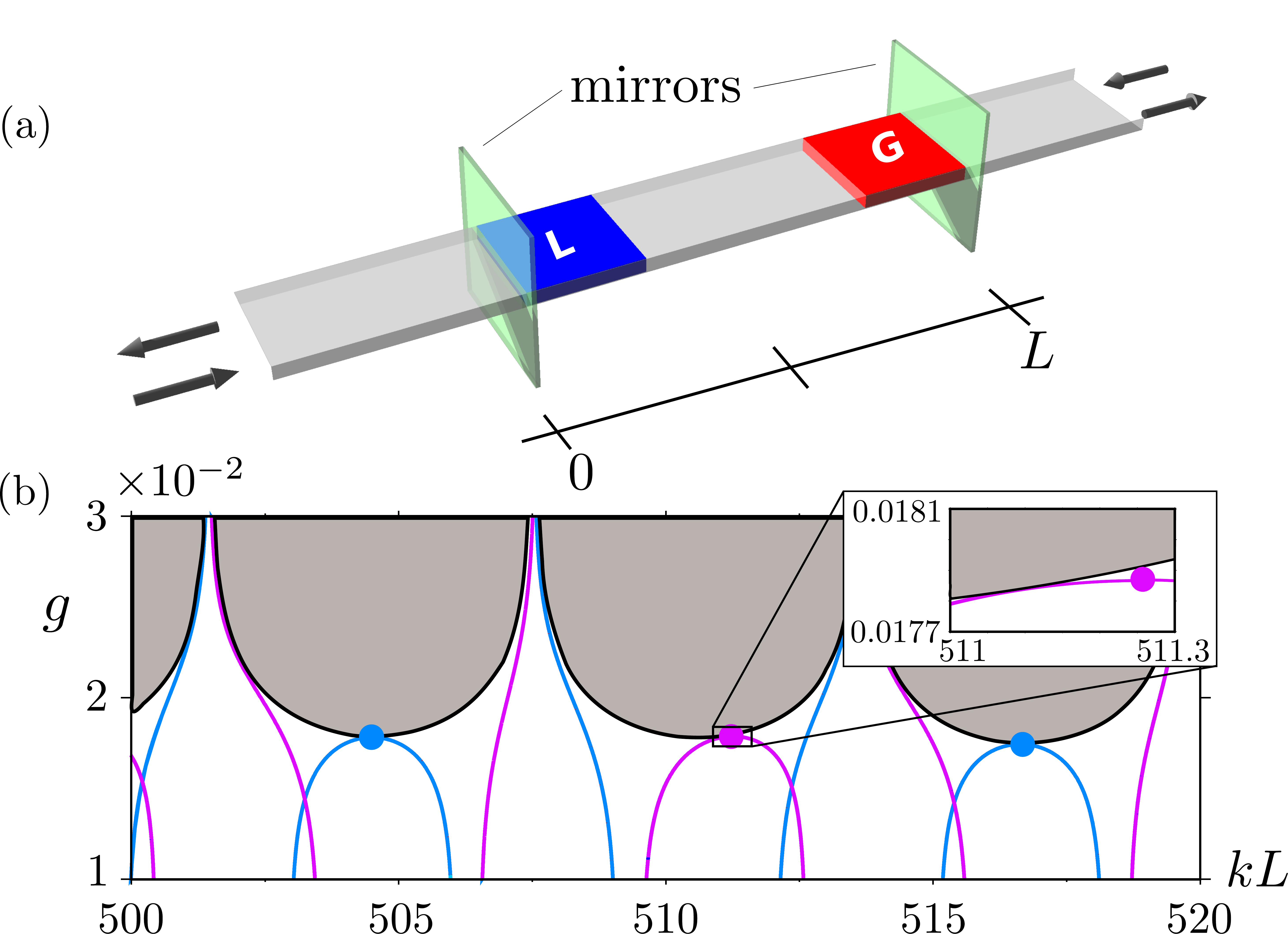}
    \caption{(a) One-dimensional scattering system featuring a layer of loss (blue) and gain (red), separated by an air gap (the real part of the refractive index $n_0=1$ throughout). A variable coupling strength between the cavity and the asymptotic region can be introduced by two semi-transparent mirrors (green). (b) Regions of unbroken (white) and broken (grey) $\mathcal{PT}$--symmetry for the scattering matrix of the system displayed in (a) ($\mathcal{P}$ denotes spatial reflection at $x\!=\!L/2$). The ``exceptional line'' (black) at the boundary between these two regions contains all the $\mathcal{PT}$--breaking points in the scattering matrix eigenvalues occurring at specific values of $kL$ and strength of loss and gain $g$. Embedding into this plot also the (real) eigenvalues below threshold of the Helmholtz equation in the bounded domain with Dirichlet (blue) or Neumann (purple) BC imposed at $x=0,L$, we find that the symmetry breaking points of these eigenvalues (see colored dots) lie in close vicinity but not exactly on the exceptional line of the $S$-matrix eigenvalues (see inset).}
    \label{fig1}
  \end{center}
\end{figure}

Consider, as a starting point, the one-dimensional (1D) $\mathcal{PT}$--symmetric scatterer shown in Fig.~\ref{fig1}(a), which is similar to those used in previous studies \cite{ChoLiSto2011,ChoLiSto2011,LieGeCer2012}. This scattering system consists of a cavity, to which two semi-infinite leads of uniform real refractive index are attached on the left and right.  The cavity itself consists of three layers, one with loss (L), one with gain (G), and an air gap in the middle. Its refractive index satisfies the $\mathcal{PT}$--symmetry relation $n\!\left(x\right)\!=\!n^{*}\!\left( \mathcal{P} x \right)$, where in this case the parity operator $\mathcal{P}$ performs a spatial reflection about $x\!=\!L/2$. We take the refractive index to be $n=n_0+ig$ ($n=n_0-ig$) in the loss (gain) regions, and $n=1$ in the air gap and in the leads. The real parameter $g$ controls the magnitude of the balanced gain and loss. The harmonic electric field $E$ transverse to the propagation axis satisfies the Helmholtz equation
\begin{equation}\label{Helmholtz}
\left[\partial^2_x+n^2\!\left(x\right)k^{2}\right]E\!\left(x,k\right)=0,
\end{equation}
with $k = \omega/c$; solutions to this unbounded problem exist for all $k\!\in\!\mathbb{R}$.
In the leads, $E$ can be expanded into incoming and outgoing waves:
\begin{equation}\label{1D-scattering-state}
E\!\left(x,k\right)=\begin{cases}
v_{1}\,e^{i k x}+u_{1}\,e^{-i k x}, & x\leq 0\\
v_{2}\,e^{-i k \left(x-L\right)}+u_{2}\,e^{i k \left(x-L\right)}, & x\geq L,
\end{cases}
\end{equation}
where $L$ is the length of the cavity. The wave amplitudes $\{\vec{u},\vec{v}\}$ are related by the scattering matrix, $\vec{u}\!=\!S\,\vec{v}$.
As noted, one can show \cite{Scho2010,ChoLiSto2011}
that either the eigenvalues of $S$ are unimodular, in which case each eigenvector $\vec{v}_j$ is $\mathcal{PT}$--symmetric ($\mathcal{P}\vec{v}_j^{\,*} \propto \vec{v}_j$), or the eigenvalues are inverse conjugates, in which case the eigenvectors break $\mathcal{PT}$--symmetry ($\mathcal{P} \vec{v}_1^{\,*} \propto \vec{v}_2$).
The $\mathcal{PT}$--phase diagram (i.e., the $kL-g$ parameter space) of the scatterer is shown in Fig.~\ref{fig1}(b).  The $\mathcal{PT}$--symmetric part of the diagram is shown in white, and the $\mathcal{PT}$--broken part is shown in grey.  These two regions are separated by an \textit{exceptional line} (black), which marks the threshold of the $\mathcal{PT}$--breaking transition; everywhere on this line, the $S$ matrix is defective (its eigenvectors are linearly dependent and it cannot be diagonalized) \cite{ChoLiSto2011,Hei2012}. Strictly speaking, due to dispersion, exact $\mathcal{PT}$--symmetry cannot hold as $k$ is varied over an interval \cite{ZyaVinDor2012}, hence we imagine varying the parameters $L,g$ \cc{to probe the $\cal PT$-breaking transition} \cite{MakAmbGe2013}.

Our goal is to relate the physics of the unbounded scattering problem to that of a bounded system with the same complex refractive index, $n(x)$, defined in the finite domain $x\!\in\!\left[0,L\right]$, and appropriate BC at $x=0,L$. In case this BC is itself $\mathcal{PT}$--symmetric, the corresponding $\mathcal{PT}$--symmetric bounded problem will undergo a transition
in which its discrete eigenvalues $k_m$ will go from being real, to being complex conjugate pairs.
The work of Ref.~\cite{GeChoSto2012} looked at the case of Dirichlet BCs, and found a rough correspondence, for a given value of $g$, between the points at which the $k_m$ are real and the intervals over which the $\mathcal{PT}$--symmetry of $S$-matrix in the unbounded problem is unbroken. In Fig.~\ref{fig1}(b) we plot the trajectories of the real eigenvalues $k_m$ as a function of $g$ for both the case of Dirichlet and of Neumann BC; as is well-known, pairs of such eigenvalues eventually meet at exceptional points of the bounded problem as the  $\mathcal{PT}$--transition occurs (we don't plot them after they  become complex). While the exceptional points of these specific bounded problems occur {\it near} the exceptional lines of the unbounded problem, close inspection shows that they do not typically occur {\it on} the exceptional lines. We now show that a more subtle relationship exists between the bounded and the unbounded problem.

Consider an incident pair of waves $\vec{v}$ corresponding to an eigenvector of the $S$ matrix, denoted $\vec{v}_i$, with complex eigenvalue $\sigma_i$:
\begin{equation}\label{S-eigenstates}
S
\left(\begin{array}{c}
v_{i,1}\\
v_{i,2}
\end{array}\right)
=
\left(\begin{array}{c}
u_{i,1}\\
u_{i,2}
\end{array}\right)
=
\sigma_{i}
\left(\begin{array}{c}
v_{i,1}\\
v_{i,2}
\end{array}\right).
\end{equation}
Let $\psi_i(x)$ denote the corresponding field, obtained by inserting the coefficients of $\vec{v}_i$ and $\vec{u}_i$ into Eq.~(\ref{1D-scattering-state}) and by solving Eq.~(\ref{Helmholtz}).  At the boundaries $x\!=\!0,L$, this scattering eigenfunction must obey
\begin{eqnarray}
\psi_{i}\!\left(0\right)\!=\!v_{i,1}\left(1+\sigma_{i}\right), & \; -\partial_{x}\psi_{i}\!\left(0\right) & =-v_{i,1}\,ik\!\left(1-\sigma_{i}\right)\label{psiLatB}\\
\psi_{i}\!\left(L\right)\!=\!v_{i,2}\left(1+\sigma_{i}\right), & \;  \partial_{x}\psi_{i}\!\left(L\right) & =-v_{i,2}\,ik\!\left(1-\sigma_{i}\right)\label{psiRatB}.
\end{eqnarray}
We now ask what choice of BC a \textit{bounded} system must have in order to possess an eigenfrequency $k_m$ which is equal to the value of $k$ chosen for the unbounded system. Note that for this BC also the eigenfunction $E_m(x)$ coincides with the scattering field $\psi_i(x)$ within the scatterer, $x \in [0, L]$. It is easy to see that Eqs.~(\ref{psiLatB})-(\ref{psiRatB}) define Robin BCs, of the form
\begin{equation}\label{RobinBC}
-\partial_{x}\psi_{i}\!\left(0\right)  =  \lambda\,\psi_{i}\!\left(0\right), \quad \partial_{x}\psi_{i}\!\left(L\right)  =  \lambda\,\psi_{i}\!\left(L\right),
\end{equation}
where the Robin parameter $\lambda$  is related to $\sigma_i$ by
\begin{equation}\label{sigmaoflambda}
  \lambda_i\!\left(\sigma_i\right)=ik\,\frac{\sigma_i-1}{\sigma_i+1}\quad
  \Longleftrightarrow\quad
  \sigma_i\!\left(\lambda_i\right)=\frac{ik+\lambda_i}{ik-\lambda_i}\,.
\end{equation}

We have thus found, by construction, an exact mapping between one of the scattering eigenstates of the unbounded system, at \cc{any arbitrary} wavevector $k$ and gain-loss parameter $g$, and a particular eigenstate of a specific bounded system with Robin BC  \cc{whose eigenfrequency equals $k$ at the same value of $g$}.
Since the $S$ matrix depends parametrically on $k$ and $g$, the relevant bounded system has a different BC at each point in the phase diagram and a \cc{\it real} eigenfrequency $k_m = k$. Furthermore, the construction also works in the opposite direction: given any bounded problem obeying the BC (\ref{RobinBC}), and a choice of any one of its real eigenfrequencies $k_m$, the corresponding scattering matrix $S(k_m,g)$ \textit{must} have an eigenvalue given by the right hand equation in (\ref{sigmaoflambda}). This mapping is completely general for any 1D Helmholtz equation and $n(x)$; we did not use Hermiticity or $\cal{PT}$--symmetry in deriving Eq.~(\ref{sigmaoflambda}).
A transformation very similar to Eq.~(\ref{sigmaoflambda}) is, in fact, used extensively in microwave engineering and known there under the name of  ``Smith charts'' \cite{Smi1939,muller_3-d_2011}. This concept maps the normalized impedance $z$ of a one-port system (one input, one output port) to its reflection coefficient $\rho$ through a M{\"o}bius transformation, $\rho\!=\!(z-1)/(z+1)$. Rewriting (\ref{sigmaoflambda}) as $-\sigma_i\!=\!(i\lambda/k-1)/(i\lambda/k+1)$, we immediately see the equivalence of the two transformations if we interpret $-\sigma_i$ as $\rho$ (the minus sign is due to our convention for the $S$ matrix) and $i\lambda/k$ as $z$. Note, however, that differently from the conventional concept of Smith charts, our approach from above applies to systems with an arbitrary number of ports. What we thus find is that a subdivision of a multi-port scattering problem into independent scattering matrix eigenchannels allows us to assign to each of these channels its own single-port M{\"o}bius transformation and with it a corresponding Smith chart. We speculate that this approach might also find applications in multi-port microwave scattering problems.

\begin{figure}
  \begin{center}
    \hspace*{-0.5cm}
    \includegraphics[angle=0, scale=0.14]{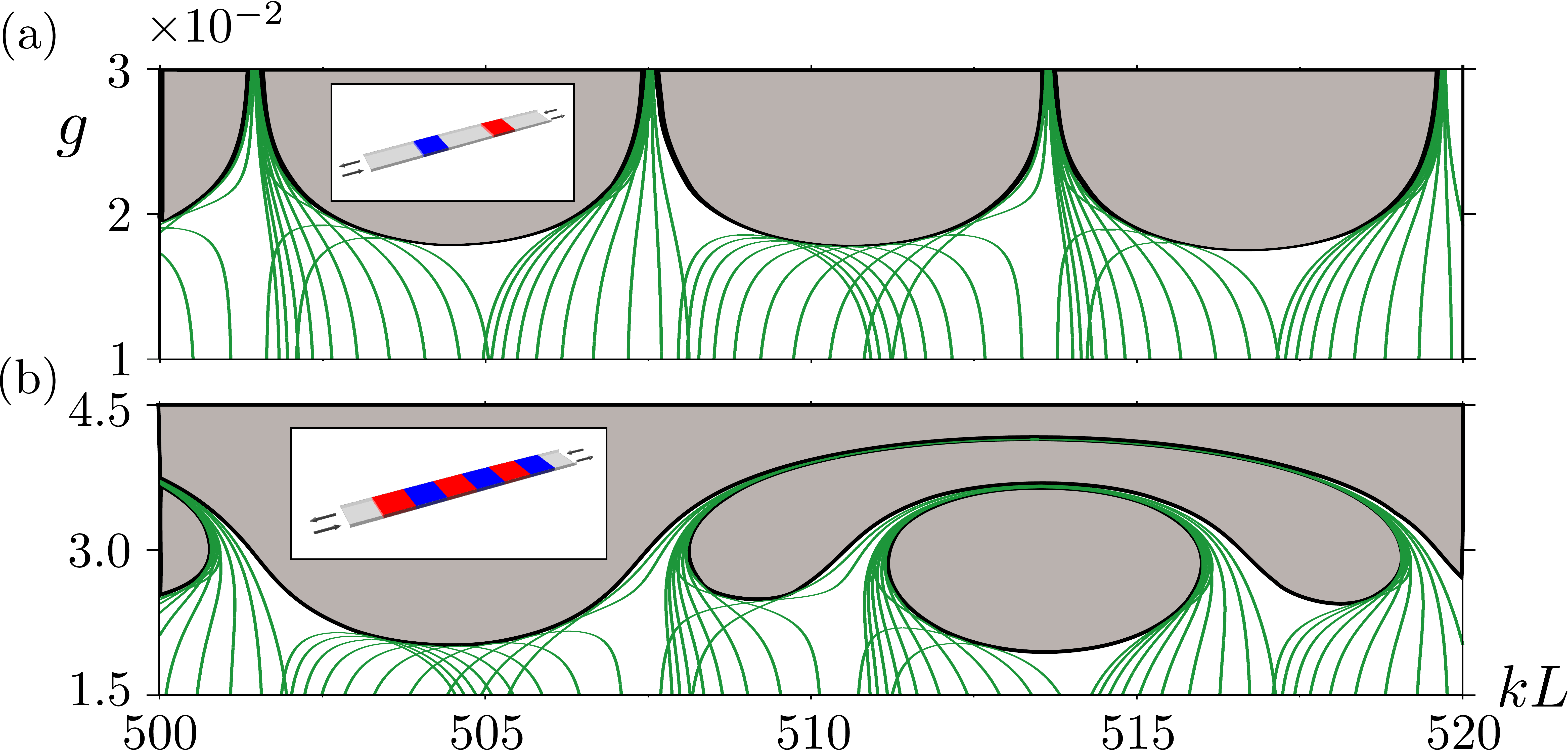}
    \caption{Eigenvalues $k_n$ (green) of the Helmholtz equation, Eq.~(\ref{Helmholtz}), in the bounded domain with Robin BC as defined in Eq.~(\ref{sigmaoflambda}). Only the real eigenvalues below the $\mathcal{PT}$--threshold are shown as a function of the gain/loss-parameter $g$. (Altogether ten different values of the boundary parameter $\lambda\!\in\!\left[-500,+500\right]$ were used.) The envelope of the eigenvalues in this bounded system corresponds exactly to the exceptional line (black), which separates the unbroken (white) from the broken (grey) $\mathcal{PT}$--phase in the corresponding unbounded scattering problem. Results are shown for (a) a two-layer setup as displayed in Fig.\ref{fig1}(a) as well as for (b) a more complicated system featuring altogether six layers of loss and gain (see insets).}
    \label{fig2}
  \end{center}
\end{figure}

For the $\mathcal{PT}$--symmetric case considered here, the mapping (\ref{sigmaoflambda}) has the following immediate implications: In the $\mathcal{PT}$--symmetric phase of the $S$ matrix, both $\sigma_i$'s are unimodular, and the $\lambda_i$'s are real. In this case the M{\"o}bius transformation in Eq.~(\ref{sigmaoflambda}) maps values of $\sigma$ on the complex unit circle onto the entire real $\lambda$ axis and vice versa.  With real $\lambda_i$, the Robin BC are Hermitian and satisfy $\mathcal{P}$-- and $\mathcal{T}$--symmetry separately, \cc{even though the heterostructure itself does not}. It follows that the union of \cc{the trajectories of all {\it real} eigenvalues $k_m$} in the {\it bounded}  $\mathcal{PT}$--problem coincide with the unimodular phase of $S$ in the $kL-g$ plane, as the real $\lambda$ varies from $-\infty$ to $\infty$ and $g$ varies from zero to $\infty$.  Thus there is no simple correspondence between the $\mathcal{PT}$--transition in scattering and any specific bounded problem; each bounded problem, however, contains information about the $\mathcal{PT}$--phase diagram. This statement is illustrated in Fig.~\ref{fig2}(a),(b) in which we vary \cc{$g$} up to the \cc{transition point of the bounded system} for many different \cc{real} values of $\lambda$, and for two different $\mathcal{PT}$--structures of increasing complexity.

\cc{Note that the above statement does not imply that the phase boundary of the $\cal PT$ scattering problem} (every point of which is an exceptional point of $S$) coincides with the union of all exceptional points of the bounded problem [see the inset in Fig.~\ref{fig1}(b)]. \cc{Since the latter points cannot occur in the $\mathcal{PT}$--broken phase of $S$, they are, however, enclosed by the phase boundary of $S$.}
In the $\mathcal{PT}$--broken phase of $S$ the
$\sigma_i$ have left the unit circle and \cc{the two corresponding $\lambda_i$} form a complex conjugate pair at each point in the $kL-g$ plane.
Whereas the BC are then non-Hermitian and non-$\mathcal{PT}$--symmetric, the equivalence between the bounded and the unbounded quantities, $k_m=k$\cc{,} $E_m\!\left(x\right)\!=\!\psi_i\!\left(x\right)$, still holds and may be visualized using the concept of 3D Smith charts recently introduced in \cite{muller_3-d_2011}. Note how, in this way, the bounded--unbounded mapping from above
provides important information on the boundary between phases where the scattering matrix
features eigenvalues on or away from the unit circle, respectively. The more trivial cases are realized for hermitian systems,
where the scattering matrix eigenvalues are always on the unit circle, or for systems with only gain or only loss, where these
eigenvalues never fall on the unit circle.

\begin{figure}
    \begin{center}
   \includegraphics[angle=0,width=0.98\columnwidth]{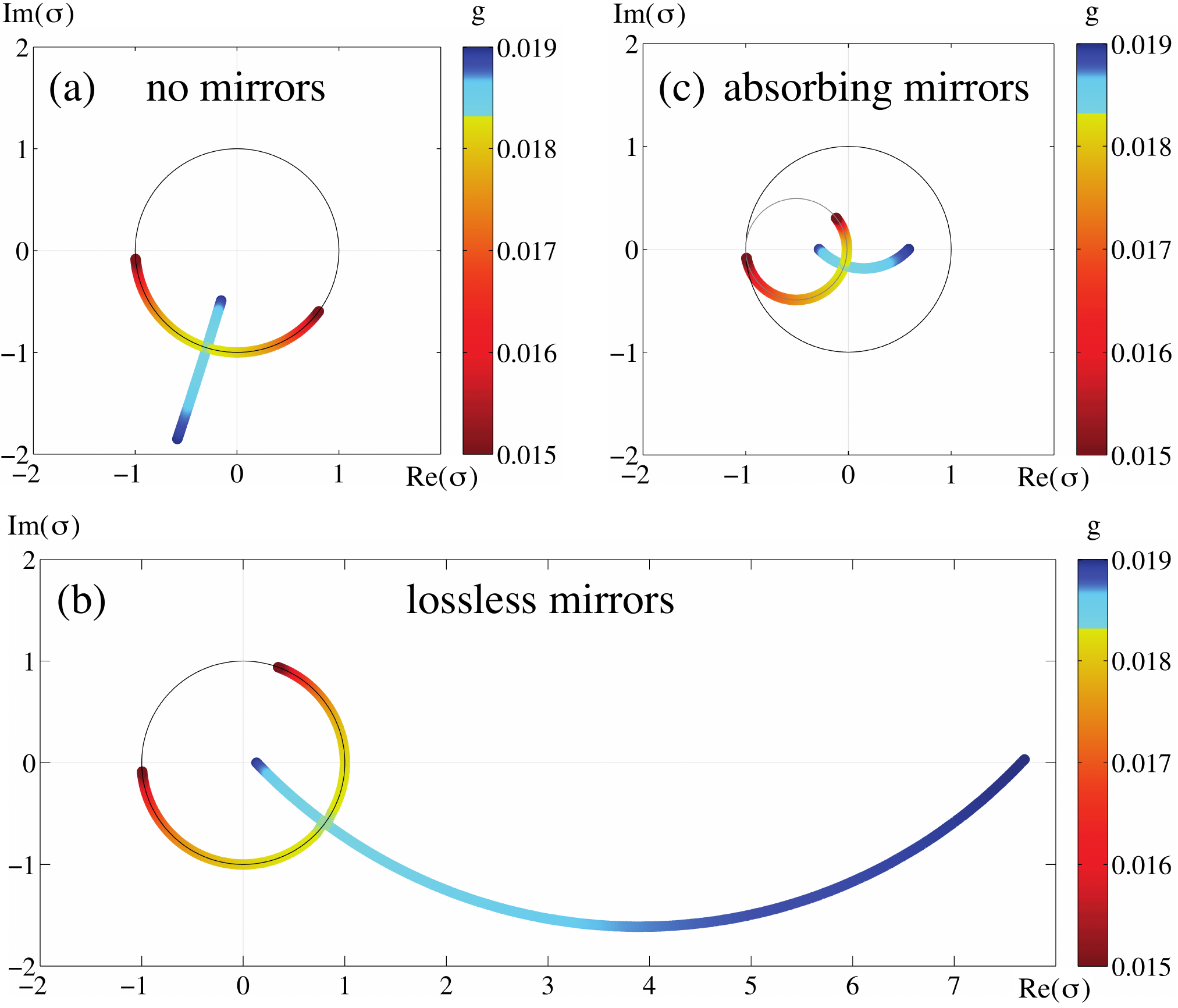}
    \caption{Movement of the scattering matrix eigenvalues $\sigma_i$ for the $\mathcal{PT}$--scattering system shown in Fig.~\ref{fig1}(a) as a function of the gain-loss strength $g$ (see color bar on the right of the panels) and at $k=515.0$. Panel (a) on the top left shows the case without externally placed mirrors where two $\sigma_i$ move on the unit circle and coincide at a critical value of $g$ (see color bar on the side). Panel (b) on the bottom shows that placing thin dielectric mirrors symmetrically around the gain-loss region rotates the $\sigma_i$ on the unit circle (below threshold) but leaves the critical gain-loss strength at which they coalesce invariant. In panel (c) on the top right we have added the same degree of absorption to both mirrors which shrinks the circle on which the $\sigma_i$ rotate (below threshold) to a smaller circle which touches the unit circle at $\sigma=-1$ and which has its center on the real axis. Also in this case the critical gain-loss strength at which the two $\sigma_i$ coalesce is the same as in (a) and(b). }\label{fig3a}
  \end{center}
\end{figure}

The mapping between bounded and unbounded problems suggests a further, previously unknown property of the $\mathcal{PT}$--transition in scattering. 
This property can be uncovered by observing that adding thin regions of real index of refraction (like a delta-function) to the scattering region symmetrically at each end (which preserves $\mathcal{PT}$--symmetry) does not change the scattering matrix eigenstates inside the mirrors 
(apart from a global amplitude). Instead, the mirrors just shift
the boundary parameter $\lambda$ of the corresponding bounded problem without mirrors by a real value,  $\lambda\rightarrow\bar{\lambda}\!=\!\lambda\!+\!\mu,\;\mu\in\mathbb{R}$ (where $\mu$ just depends on the reflectivity of the mirror). Changing the BC in this sense, however, does not change the location of the exceptional line \cc{since the $\cal PT$--symmetric phase of $S$ is the union of {\it all} real values of $\lambda$ as mentioned, from $-\infty$ to $\infty$}. This result can be conveniently visualized with a Smith chart (see Fig.~\ref{fig3a}), where the shift of $\lambda$ can be seen to just rotate both scattering matrix eigenvalues on the unit circle, which leaves the gain/loss-strength $g$ at which they coalesce invariant [compare Fig.~\ref{fig3a}(a) and (b)]. We thus arrive at the conjecture that many different $\mathcal{PT}$--scattering problems have the same $\mathcal{PT}$--phase diagram, if they differ only by the addition of dielectric ``mirrors" at the two ends. This conjecture can be proved rigorously in 1D for arbitrary lossless dielectric structures added to the original $\mathcal{PT}$--cavities (see appendix~\ref{appA} for this proof which also holds for thick dielectric structures featuring several dielectric layers).
We have tested this ``mirror theorem" also numerically by adding lossless
mirrors [see Fig.~\ref{fig1}(a)] to the ends of the six-layer scattering system in Fig.~\ref{fig2}(b), and find that its complicated phase boundary is reproduced to within the numerical accuracy of the computation.

\cc{The above ``mirror theorem'' also features an interesting manifestation in the ratio of the incoming (outgoing) amplitudes  $\xi_i\equiv v_{i,1}/v_{i,2}(=u_{i,1}/u_{i,2})$ of the scattering eigenstates in 1D, whose modulus and phase display a bifurcation at the phase boundary of $S$, respectively \cite{GeChoSto2012}. As we show in appendix~\ref{appA}, $\xi_i$ is invariant not just at the exceptional line with the addition of lossless symmetric mirrors; it is so everywhere in the phase space, both in the $\cal PT$--symmetric phase and broken phase. More surprisingly, this property holds even when the symmetric mirrors added are dissipative or amplifying, which clearly violates the global $\cal PT$--symmetry of the heterostructure. This finding implies that the phase boundary of $S$ is also invariant in this general situation and we find, indeed, that} the scattering eigenvalues $\sigma_i$ still meet at exactly the same exceptional line as in the case without such mirrors. For thin non-hermitian mirrors with loss or gain, this situation can again be understood by the shift which these mirrors induce on the Robin BC parameter $\lambda\to\bar{\lambda}=\lambda+\mu$ with $\mu$ now being complex rather than real as before. As visualized conveniently on a Smith chart, the real part of $\mu$ leads again to a rotation of the scattering matrix eigenvalues $\sigma_i$, but its imaginary part shrinks (expands) the circle below (beyond) the unit circle on which they rotate in the $\mathcal{PT}$--unbroken phase [see Fig.~\ref{fig3a}(c)]. Both of these operations do, however, leave the critical gain/loss-strength $g$ at which the two scattering matrix eigenvalues $\sigma_i$ coalesce invariant. \cc{Since both the scattering eigenvectors and eigenvalues still coalesce at the original exceptional line}, we arrive at the very general result that this line is entirely unaffected by \cc{the symmetric} mirrors, even if they are absorbing or amplifying. In appendix \ref{appA} we provide a rigorous proof of this result even for thick stacks of absorbing or amplifying mirrors and check this result also explicitly numerically \cc{(see Fig.~\ref{fig4}). The generalization of the mirror theorem to non-hermitian mirrors is particularly important for two reasons: First, it shows that the exceptional points in a $\cal PT$--symmetric system are not necessarily a result of the global $\cal PT$-symmetry; they persist even when the symmetry is broken when the absorptive or amplifying mirrors are added. Second, it paves the way for the experimental verification of the ``mirror theorem,'' since in practice the mirrors can never be absolutely loss-free.}


\begin{figure}
    \begin{center}
   \includegraphics[width=\linewidth]{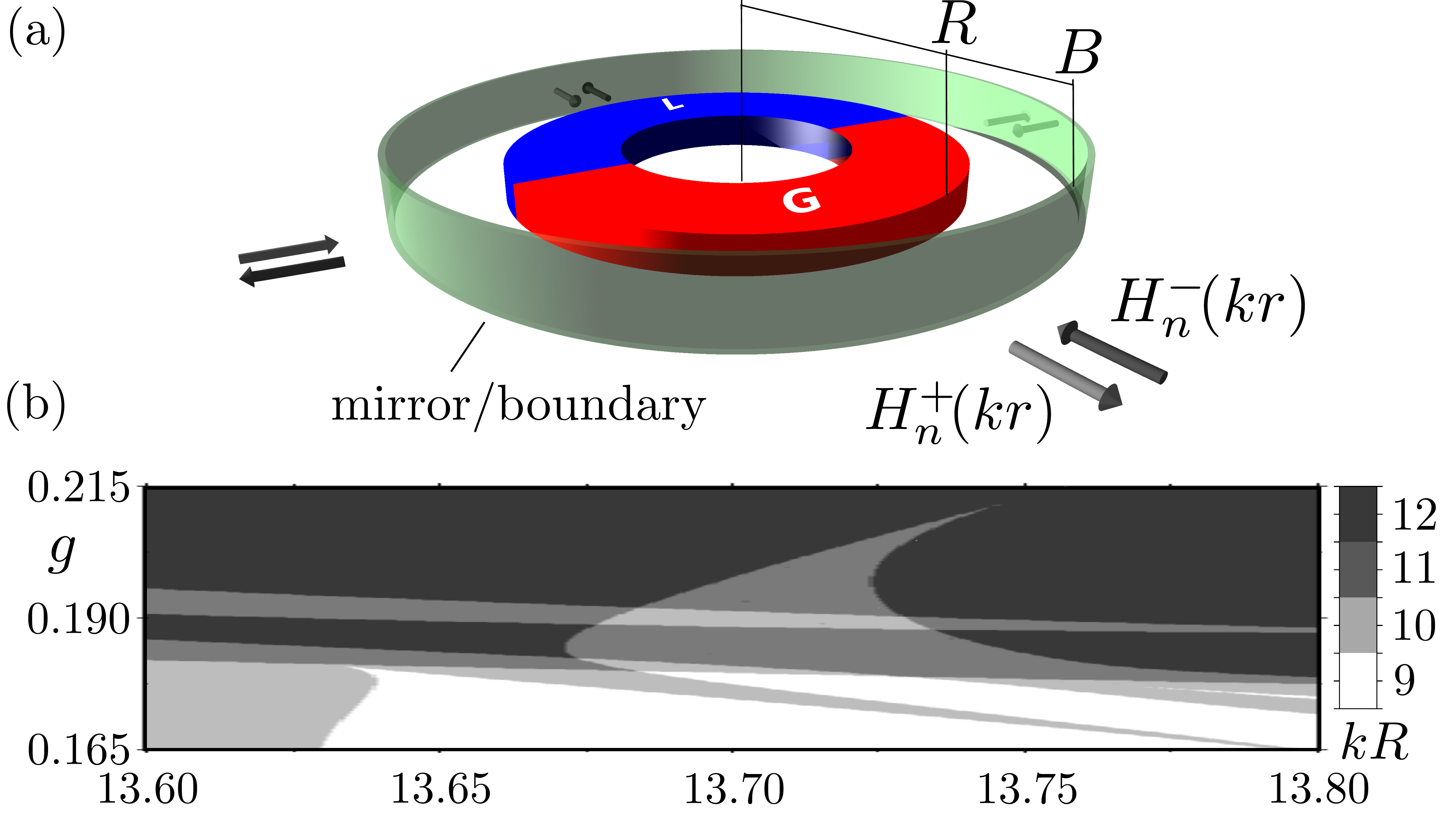}
    \caption{(a) $\mathcal{PT}$--symmetric disk of radius $R$ consisting of one region with loss (blue) and one with gain (red), enclosing an air gap ($n_0=1$ throughout). As in 1D, a partially transparent mirror (green) can be used to vary the coupling strength to the asymptotic region. (b) A small part of the complicated phase diagram with overlapping phases of broken und unbroken $\mathcal{PT}$--symmetry for the $S$-matrix eigenstates in this system. Different shades of grey correspond to different numbers of pairs of scattering matrix eigenvalues in their respective $\mathcal{PT}$--broken phases. As we verified explicitly with our numerical simulations for this device, the phase diagram stays unchanged when tuning the reflectivity of the mirror placed in the far-field (we chose a value $B\!=\!1000\times R$).}\label{fig3}
  \end{center}
\end{figure}

Our mapping approach to connecting bounded and unbounded scattering problems suggests that some form of the mirror theorem
could also hold in higher dimensions; we will now demonstrate a two-dimensional example of this (some general relations that hold 
for arbitrary 2D $\mathcal{PT}$--symmetric scattering problems are provided in appendix \ref{appD}). 
Consider the case of a two-dimensional $\mathcal{PT}$--symmetric disk of radius $R$ as shown in Fig.~\ref{fig3}(a).
To evaluate a scattering matrix for such a system, we envision a circular boundary with radius 
$B>R$ [see Fig.~\ref{fig3}(a)] outside of which we define an
appropriate scattering basis as the product of normalized incoming ($-$) and outgoing ($+$) Hankel functions, $H^\pm_{n}\!\left(k r\right)\!\equiv\cc{{\rm H}}^\pm_{n}\!\left(k r\right)\!/\cc{{\rm H}}^\pm_{n}\!\left(k B\right)$, and normalized trigonometric functions $\chi_{n,1}\!\left(\varphi\right)\!=\!A_1\sin\!\left(n\varphi\right),\,\chi_{n,2}\!\left(\varphi\right)\!=\!A_2\cos\!\left(n\varphi\right)$.
For the infinite dimensional scattering matrix defined in this basis there are typically many eigenvalue pairs which
go through a $\mathcal{PT}$--transition, leading to a very complicated phase diagram [a small detail of which is shown in Fig.~\ref{fig3}(b)].
Still, we can represent the scattering eigenstates in the bounded domain $r<B$ as the eigenstates of a boundary value problem (with the boundary
at $r\!=\!B$). For this purpose we first expand the scattering matrix eigenstates $S\vec{v}_i=\sigma_i\vec{v}_i$ for $r>B$ in the above basis $\psi_i=\sum_{n,\eta} v^i_{n,\eta} \chi_{n,\eta}(H^{-}_{n}+\sigma_i H^+_{n})$. Making a Robin-ansatz for the boundary conditions of these states $\psi_i$ (as in 1D),
\begin{equation}
  \partial_r\psi_{i}\!\left(\vec{x}\right)
  \!=\!\lambda_{i}\psi_{i}\!\left(\vec{x}\right),\quad
  \lambda_{i}\!\equiv\!\lambda\!\left(\sigma_{i},\varphi\right).\label{lambda-gen}
\end{equation}
we find that the corresponding factors $\lambda_i$ appearing here do, in general, not just depend on the eigenvalue $\sigma_{i}$ but also on the angular position $\varphi$ on the boundary. The resulting angle-dependent Robin boundary condition takes the following form,
\begin{equation}
  \partial_r \psi_{i}(r,\varphi)|_{r\!=\!B}
  \!=\!\sum_{n,\eta}\Lambda^i_{n}\,
  \chi_{n,\eta}\!\left(\varphi\right)\int_{0}^{2\pi}\! d\theta\,\chi_{n,\eta}\!\left(\theta\right)\psi_{i}\!\left(B,\theta\right)\,,\label{lambda-modes}
\end{equation}
with coefficients $\Lambda^i_{n}\!=\!\left(\partial_r H^-_{n}+\sigma_{i} \,{\partial_r H^{+}_{n}}\right)|_{r\!=\!B}\left(1+\sigma_{i}\right)^{-1}$. These coefficients do, however, lose their $n$-dependence if we choose the boundary of our finite domain in the far-field, i.e., $B\gg R$.
In this case the Hankel-functions can be approximated by $n$-independent cylindrical waves,
$H^\pm_{n}\!\left(k r\right)\approx e^{\pm ik\left(r-B\right)}/\sqrt{r/B}$ for $r>B$ and the expressions in
Eq.~(\ref{lambda-modes}) drastically simplify. As a result, the terms $\lambda_i$ in the Robin
boundary condition, Eq.~(\ref{lambda-gen}), are then given by the coefficients $\Lambda^i_{n}$ which are independent
of $n$ and thus of $\varphi$. Neglecting contributions of lower order than $r^{-1/2}$ we find,
\begin{equation}
  \lambda\!\left(\sigma_{i},\varphi\right)\equiv\lambda_{i}= ik\,\frac{\sigma_i-1}{\sigma_i+1}=const.,
\end{equation}
which is exactly the same expression, Eq.~(\ref{sigmaoflambda}), which we have previously obtained in 1D.
In the same way as we have argued in 1D that a mirror placed symmetrically around the $\mathcal{PT}$--system does not change the exceptional line, we can now make the same conjecture for \textit{each individual} eigenvalue of a 2D scattering matrix evaluated in the far-field. Hence the complicated $\mathcal{PT}$--phase diagrams as in Fig.~\ref{fig3}(b) should not change with the addition of concentric mirrors in the far-field.  Again we confirm this conjecture by numerical tests involving the structure shown in Fig.~\ref{fig3}(a).

The mirror theorem indicates that the $\mathcal{PT}$--transition in scattering is quite a subtle phenomenon. If we think of the $\mathcal{PT}$--symmetric scattering region as a resonator, adding (non-absorbing) external mirrors greatly enhances the $Q$ value (i.e., the cavity lifetime) of such a resonator, but apparently has no effect on its phase boundary. The reasoning that having the waves stay in the resonator much longer would allow them to feel the presence of gain and loss more strongly and would thus lead to a $\mathcal{PT}$--transition at smaller $g$ therefore proves incorrect. Our results thus dramatically illustrate that the $\mathcal{PT}$--breaking transition in scattering is not a resonance phenomenon and that it does not depend on quantities like the round-trip gain/loss that are used to estimate the lasing transition.
Instead, the $\mathcal{PT}$--transition is sensitive to the coupling of the gain and loss regions with each other, with strong coupling making the transition harder to achieve and weak coupling making it (trivially) easier. Quite on the contrary, the coupling to the external world has no effect at all on the transition, but strongly affects other features of the resonator which are sensitive to higher $Q$-values. Consider, e.g., those singular points in the broken symmetry phase \cite{Mos2009,Lon2010,ChoLiSto2011,Scho2010,YooSimScho2011,SchiLinLee2012} at which one eigenvalue of the 1D $S$-matrix goes to infinity, corresponding to the laser threshold, and the other one goes to zero, corresponding to coherent perfect absorption (CPA) \cite{ChoLiSto2011}. If one adds highly reflecting dielectric mirrors to a simple low-$Q$ $\mathcal{PT}$--resonator as in Fig.~\ref{fig1}(a), these singular CPA-Laser points are pulled down almost to the $\mathcal{PT}$-phase boundary which itself, however, stays unchanged. This behavior, details of which are shown in Fig.~\ref{fig5} of appendix~\ref{appC}, nicely illustrates that $\mathcal{PT}$--symmetry breaking and the lasing transition are very different phenomena.

In summary, we have uncovered a close link between the phase transitions in the scattering matrix $S$ of an unbounded $\mathcal{PT}$--symmetric system and the corresponding transitions in the underlying bounded systems. The most interesting result which follows from this relation is the fact that under very general conditions the $\mathcal{PT}$--thresholds in the scattering matrix are unchanged by external mirrors which increase the $Q$ values of the scattering system. This prediction should be directly testable in the $\mathcal{PT}$--symmetric scattering experiments that have recently been realized.

We would like to thank S.~Burkhardt, R.~El-Ganainy, U.~G\"unther, and M.~Liertzer for helpful
discussions. Financial support by the following funding sources is gratefully acknowledged:
Vienna Science and Technology Fund (WWTF) through Project No.~MA09-030
(LICOTOLI); Austrian Science Fund (FWF) through Project No.~F25-P14 (SFB IR-ON) and
No.~F49-P10 (SFB NextLite); People Programme (Marie Curie Actions) of
the European Union's Seventh Framework Programme (FP7/2007-2013) under
REA grant agreement number PIOF-GA-2011-303228 (project NOLACOME);
Singapore National Research Foundation under grant No.~NRFF2012-02;
US National Science Foundation under grant No.~ ECCS 1068642.
We are also grateful for free access to the computational resources of the
Vienna Scientific Cluster (VSC).

\def\ket#1{\left|#1\right\rangle}
\def\bra#1{\left\langle#1\right|}
\def\braket#1{\left\langle#1\right\rangle}
\newcommand{\gper}{\gamma_\perp}
\newcommand{\gpar}{\gamma_\parallel}
\newcommand{\be}{\begin{equation}}
\newcommand{\ee}{\end{equation}}
\newcommand{\bea}{\begin{align}}
\newcommand{\eea}{\end{align}}
\newcommand{\bx}{\bm{x}}
\newcommand{\pt}{$\cal PT$}

\begin{appendix}

\section{Mirror Theorem for \pt--phase transitions in scattering through 1D heterostructures \label{appA}}

In the main text we argue that \cc{the exceptional line} of the scattering matrix \cc{and the incoming (outgoing) amplitude ratio $\xi$ of scattering eigenstates} are strictly independent of the coupling strength to the asymptotic regions if \cc{symmetric} mirrors are added at both boundaries of the \cc{1D} scattering system. Here we prove this independence explicitly, \cc{whether or not the mirrors are lossless.}

The 1D scattering matrix $S$ connecting incoming ($\vec{v}$) with outgoing ($\vec{u}$) coefficients is defined by
\be
\begin{pmatrix}
u_1 \\ u_2
\end{pmatrix}
= S
\begin{pmatrix}
v_1 \\ v_2
\end{pmatrix}
\equiv
\begin{pmatrix}
r_L & t \\ t & r_R
\end{pmatrix}
\begin{pmatrix}
v_1 \\ v_2
\end{pmatrix},
\ee
where $r_{L(R)}$ denote the reflection coefficients for injection from the left (right) and $t$ the transmission amplitude. It is easy to see, that its eigenvalues \cc{and the incoming (outgoing) amplitude ratios $\xi\equiv v_{1}/v_{2}(=u_{1}/u_{2})$} of scattering matrix eigenstates are given by
\begin{align}
  \sigma_{1,2}&=\frac{r_L+r_R}{2}\pm\sqrt{\frac{\left(r_L-r_R\right)^2}{4}+t^2},\label{evS} \\
  \xi_{1,2}&=\frac{r_L-r_R}{2t}\pm\frac{1}{t}\sqrt{\frac{\left(r_L-r_R\right)^2}{4}+t^2},\label{eveS}
\end{align}
respectively. Along the exceptional line, the two eigenvalues \cc{and eigenvectors} coalesce, i.e., the square root on the right-hand sides of (\ref{evS}),(\ref{eveS}) vanish. Thus, the scattering coefficients satisfy \cite{ChoLiSto2011}
\be
Y \equiv \frac{r_L}{t}-\frac{r_R}{t}=\pm2i\label{eq:pt}
\ee
at the $\cal PT$-phase transition points, and we note that $r_L$,$r_R$ are always $\pi$ out-of-phase with $t$ in a $\cal PT$--symmetric heterostructure.
Note that $Y$ is the sum of the two off-diagonal elements in the corresponding transfer matrix $M$ that connects the coefficients corresponding to the asymptotic regions to the right and to the left of the heterostructure. Accordingly, $M$ is defined by
\be
\begin{pmatrix}
u_1 \\ v_1
\end{pmatrix}
= M
\begin{pmatrix}
v_2 \\ u_2
\end{pmatrix}
=
\begin{pmatrix}
t-r_Lr_R/t & r_L/t \\ -r_R/t & 1/t
\end{pmatrix}
\begin{pmatrix}
v_2 \\ u_2
\end{pmatrix}.
\ee

We now vary the coupling of the $\cal PT$--symmetric heterostructure to the unbounded asymptotic regions by introducing two \cc{symmetric} mirrors of arbitrary complexity and extension (one on each side). The refractive indices $n_L,n_R$ of the left and right mirrors satisfy $n_L(x)=n_R({\cal P}x)$. We refer to the transfer matrices of the left mirror, of the original heterostructure, and of the right mirror, as $M_L$, $M$, and $M_R$, respectively. The total transfer matrix of the new structure is then $M'=M_LMM_R$.

Using the $\cal P$--symmetry of $n_L$ and $n_R$, we find that the transfer matrices of the mirrors are connected by $M_L = P M_R^{-1} P$ where $P$ is the matrix representation of the parity operator $\cal P$, given by
\be
P=\begin{pmatrix} 0 & 1\\ 1 & 0\end{pmatrix}.
\ee
This relation leads to
\be
M_L = \begin{pmatrix} m_{11} & -m_{21}\\ -m_{12} & m_{22}\end{pmatrix}\,,
\ee
if we write
\be
M_R = \begin{pmatrix} m_{11} & m_{12}\\ m_{21} & m_{22}\end{pmatrix}.
\ee
Further writing
\be
M = \begin{pmatrix} M_{11} & M_{12}\\ M_{21} & M_{22}\end{pmatrix}, \quad M' = \begin{pmatrix} M'_{11} & M'_{12}\\ M'_{21} & M'_{22}\end{pmatrix},
\ee
we derive
\bea
M'_{12} = &m_{11}m_{12}M_{11} + m_{11}m_{22}M_{12} \nonumber \\
&- m_{12}m_{21}M_{21} - m_{21}m_{22}M_{22} \\
M'_{21} = &-m_{11}m_{12}M_{11} - m_{12}m_{21}M_{12} \nonumber \\
&+ m_{11}m_{22}M_{21} + m_{21}m_{22}M_{22}\,.
\end{align}
\cc{Using the general property $\text{Det}(M_R) \!=\! m_{11}m_{22} - m_{12}m_{21}=1$}, we find
\bea \label{wpw}
Y' &= M'_{12}+M'_{21} \nonumber \\
&= (m_{11}m_{22} - m_{12}m_{21})(M_{12}+M_{21}) \nonumber \\
&= M_{12}+M_{21} \nonumber \\
&= Y.
\end{align}
One particular implication of this relation is that the two scattering eigenvalues $\sigma_{1,2}$ still coalesce at the original exceptional line, where $Y'\!=Y\!=\!\pm2i$.
\cc{In addition, we note that the incoming/outgoing amplitude ratios $\xi$ can be rewritten as $Y/2\pm\sqrt{Y^2/4+1}$. Thus they are also invariant in this case everywhere in the $kL-g$ plane and coalesce at the original exceptional line. Therefore, we come to the conclusion that the exceptional line is invariant with the addition of the symmetric mirrors.}


\begin{figure}[b]
    \begin{center}
      \includegraphics[angle=0,  scale=0.145]{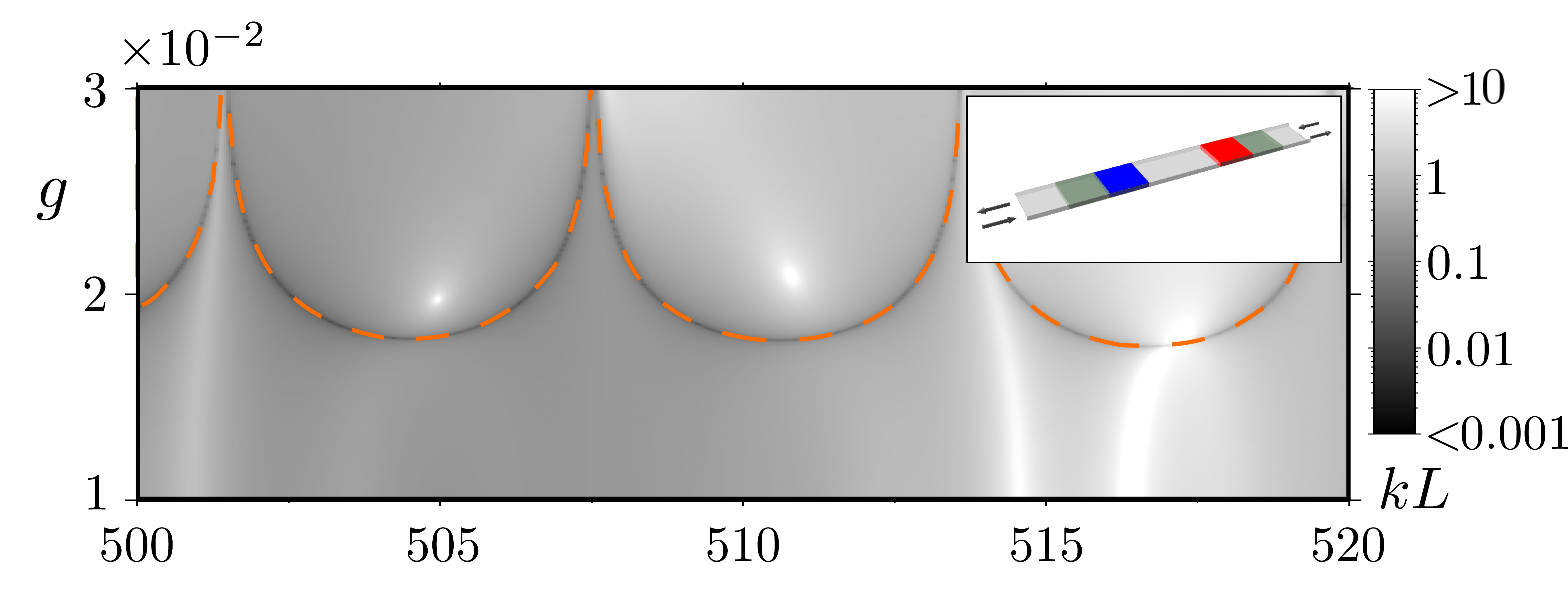}
    \caption{Difference between the two scattering matrix eigenvalues $D(kL,g)\!\equiv\!\left|\sigma_1-\sigma_2\right|$ for the  $\mathcal{PT}$--symmetric two-layer system shown in Fig.~\ref{fig1}\cc{(a)} with two slabs of width $L/4$ attached on either side (see also smaller inset). The slabs feature randomly chosen complex refractive indices that satisfy $n_R(x)=n_L({\cal P}x)$ where $R/L$ stands for for the slab on the right/left side of the $\mathcal{PT}$--symmetric structure. The lines where $D$=0, i.e., where $\sigma_1\!=\!\sigma_2$, coincide exactly with the exceptional line of the inner $\mathcal{PT}$--symmetric part of the system (orange dashed line).}\label{fig4}
  \end{center}
\end{figure}

\cc{Note that we do not impose any condition on the realness of the refractive indices $n_L,n_R$, thus the above conclusion holds for lossless mirrors, as well as for absorptive/amplifying mirrors. In the former case where $n_L,n_R$ are real, the system is still $\cal PT$--symmetric and the exceptional line is still the phase boundary of the $\cal PT$-symmetric and broken phases of $S$. In the latter case where}
the mirrors destroy the overall $\mathcal{PT}$--symmetry of the resulting system, i.e. $P S^{\dagger}PS\!\ne\!1$, the eigenvalues of the $S$-matrix generally do not lie on the unit circle and no symmetry breaking occurs. \cc{Nevertheless, the same exceptional line persists and so do the incoming/outgoing amplitude ratios $\xi$ everywhere in the $kL-g$ plane.}
To illustrate the generality of this result explicitly, we show in Fig.~\ref{fig4} the difference between the two scattering matrix eigenvalues, $D(kL,g)\equiv\left|\sigma_1-\sigma_2\right|$ for a system composed of two mirrors of width $L/4$ featuring randomly chosen complex refractive index distributions $n_L(x)$ and $n_R(x)=n_L({\cal P}x)$ attached to the two-layer system displayed in Fig.~\ref{fig1}\cc{(a)}. As can be seen by comparison to Fig.~\ref{fig1}\cc{(b)}, the union of all the points where $D\!=\!0$ is the exceptional line of the original system without mirrors.

\begin{figure}[b!]
    \begin{center}
      \includegraphics[angle=0,  scale=0.145]{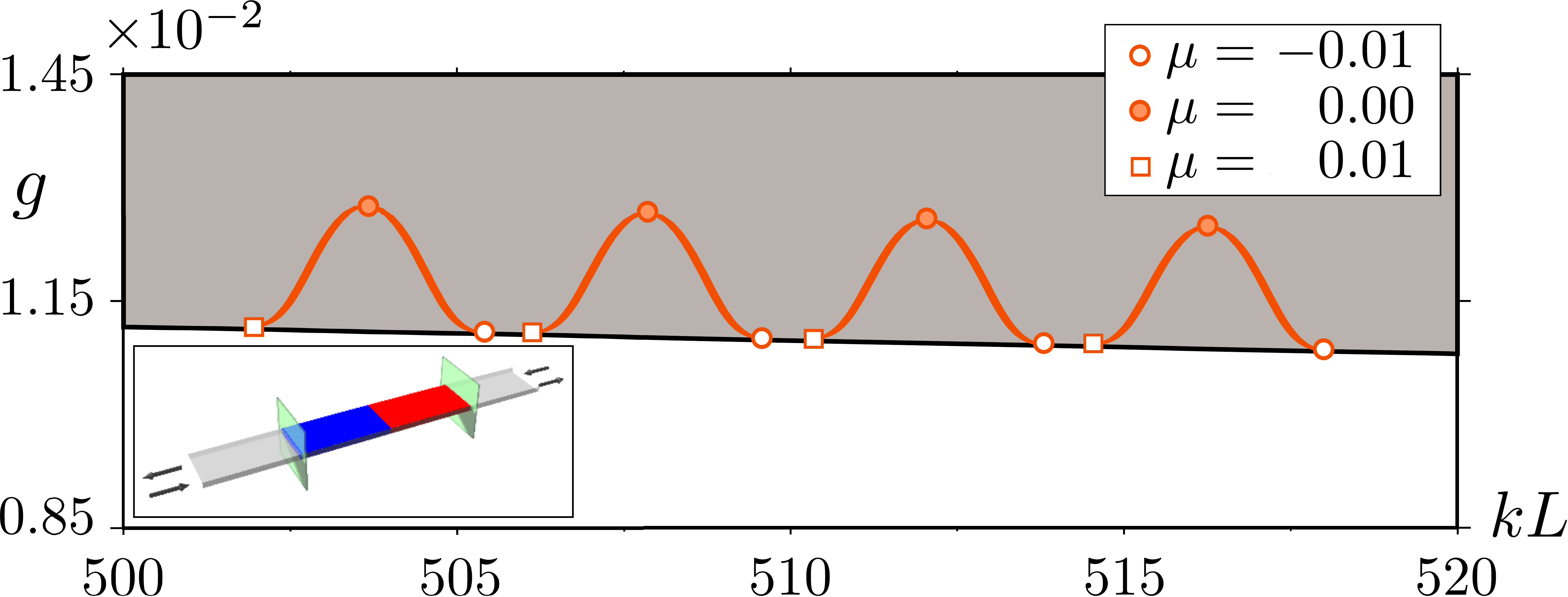}
    \caption{CPA-Laser points for a $\mathcal{PT}$--symmetric scatterer consisting of two adjacent regions of refractive index $n = 1.5 \pm ig$, with zero-width mirrors at the scattering boundary (see bottom left inset). The mirror parameter $\mu$ is defined in Eq.~(\ref{delta mirror}). The orange curves show the trajectories of four distinct CPA-Laser points for varying values of $\mu$ with the symbols indicating the positions at particular values of $\mu$ (see top right inset). The solid black curve is the exceptional line.}\label{fig5}
  \end{center}
\end{figure}

\section{Dependence of lasing transitions on the coupling strength for 1D heterostructures \label{appC}}

At discrete points in its phase space, a $\mathcal{PT}$--symmetric scatterer can act simultaneously as a laser at threshold and a coherent perfect absorber of incident light \cite{Lon2010,ChoLiSto2011}.  Since these ``CPA-Laser'' points correspond to one eigenvalue of the $S$-matrix going to zero and the other going to infinity, they can occur only within the $\mathcal{PT}$--broken phase of $S$.  Although the ``mirror theorem'' (see main text and appendix~\ref{appA}) shows that adding $\mathcal{P}$-- symmetric mirrors at the scattering boundaries does not alter the exceptional line, they do affect the position of the CPA-Laser points.

To demonstrate this invariance, we consider again a 1D $\mathcal{PT}$--symmetric scatterer (see inset of Fig.~\ref{fig5}) and modify its dielectric function by
\begin{equation}
  \Delta [n^2(x)] = \mu \left[\delta(x) + \delta(x-L)\right]
  \label{delta mirror}
\end{equation}
with a real parameter $\mu$ that controls the mirror reflectivity. This change in the dielectric function corresponds to placing dielectric mirrors of vanishing width at the scattering boundaries $x=0,L$. (The reason for using zero-width mirrors is to avoid the complication of additional internal resonances.)

Fig.~\ref{fig5} shows the trajectories of the CPA-Laser points as $\mu$ is varied, for a simple heterojunction of two equal-length gain-loss segments with refractive index $n_0 \pm ig$ (and with $n\!=\!1$ outside).  For $\mu \!=\! 0$, the CPA-Laser points are located at some distance above the exceptional line, as discussed in Ref.~\cite{ChoLiSto2011}. Increasing or decreasing $\mu$ pulls the CPA-Laser points toward the exceptional line, which can be interpreted as an effect of the increase in the $Q$-factor of the cavity.

\section {Boundary conditions for two-dimensional scattering setups}\label{appD}

In this section we provide further details on the connection between the $\mathcal{PT}$--thresholds in bounded and unbounded two-dimensional (2D) systems. Following the arguments for 1D setups, we will find that also in 2D appropriate Robin BCs can be found which give rise to eigenstates of the $S$ matrix. 

We consider an arbitrarily shaped 2D $\mathcal{PT}$--symmetric cavity through the boundary of which waves can go in and out from and to infinity.
The $m$-th eigenstate of the corresponding scattering matrix, written as a coefficient vector $\vec{v}_m$ and associated to the eigenvalue $\sigma_m$, is decomposed outside the cavity into a scattering basis $\phi_n$ as follows,
\begin{equation}
\psi_{m}\!\left(\vec{x}\right)=\sum_{n=1}^{N}v_{m,n}\left[\phi_{n}\!\left(\vec{x}\right)+\sigma_{m}\phi_{n}^{*}\!\left(\vec{x}\right)\right].\label{scattering-eigenstate}
\end{equation}
As was shown in \cite{ChoLiSto2011}, the eigenvectors of $S$ satisfy $P\mathcal{T}\vec{v}\!\propto\!\vec{v}$ below threshold and $P\mathcal{T}\vec{v}_m\!\propto\!\vec{v}_{m'}$ above threshold for an associated pair of eigenvectors ($\vec{v}_m,\vec{v}_{m'}$), respectively.
The connection between the normal derivative and the wave function of a scattering matrix eigenstate at the system boundary can 
formally be written as follows
$\psi_{m}^{\prime}\!\left(\vec{x}\right)
\!=\!
\lambda\!\left(\sigma_{m},\vec{x}\right)
\psi_{m}\!\left(\vec{x}\right)$
with the functions
\begin{equation}
\lambda\!\left(\sigma_{m},\vec{x}\right)=\frac{\sum_{n}v_{m,n}\left[\phi_{n}^{\prime}\!\left(\vec{x}\right)+\sigma_{m}\phi_{n}^{*\prime}\!\left(\vec{x}\right)\right]}{\sum_{l}v_{m,l}\left[\phi_{l}\!\left(\vec{x}\right)+\sigma_{m}\phi_{l}^{*}\!\left(\vec{x}\right)\right]}\,.\label{lambda-general}
\end{equation}
Below the $\mathcal{PT}$--threshold, where $\sigma^{*}\!=\!\sigma^{-1}$, we can rewrite the boundary-value-function $\lambda\!\left(\sigma\right)$
in vector notation and suppressing $(\vec{x}$) as follows
\begin{equation}
\lambda\!\left(\sigma\right) = 
\frac{\vec{v}^{\, T}\left(\vec{\phi}^{\prime}+\sigma\vec{\phi}^{*\prime}\right)}{\vec{v}^{\, T}\left(\vec{\phi}+\sigma\vec{\phi}^{*}\right)} =
\frac{\vec{v}^{\,\dagger}P\left(\vec{\phi}^{\prime}{}^{*}+\sigma^{*}\vec{\phi}^{\prime}\right)}{\vec{v}^{\,\dagger}P\left(\vec{\phi}^{\prime}{}^{*}+\sigma^{*}\vec{\phi}\right)} =
\mathcal{P}\lambda^{*}\left(\sigma\right),\label{lambda-rew-below}
\end{equation}
where we further used $P^{*}\!=\!P\!=\!P^{T}$ for the matrix representation $P$ of the parity-operator $\mathcal{P}$ and $\vec{v}^{\,T}\!\propto\!\vec{v}^{\,\dagger}P$. Note, that the action of $P$ and taking the normal derivative commute due to the $\mathcal{P}$--symmetry of the boundary. From (\ref{lambda-rew-below}) we see, that $\lambda\!\left(\sigma\right)$ is a $\mathcal{PT}$--symmetric function $\lambda\!\left(\sigma,\vec{x}\right)\!=\!\lambda^{*}\!\left(\sigma,\mathcal{P}\vec{x}\right)$,
leading to $\mathcal{PT}$--symmetric Robin boundary conditions for the wave function.\\
Above the $\mathcal{PT}$--threshold of an associated pair ($\sigma_{m},\sigma_{m'}$) of $S$ matrix eigenvalues, we find for the expression of $\lambda\!\left(\sigma_{m'}\right)$ using $\vec{v}_{m'}^{\,T}\!\propto\!\vec{v}_m^{\,\dagger}P$ and $\sigma_{m'}\!=\!1/\sigma_{m}^{*}$,
\begin{eqnarray}
\lambda\!\left(\sigma_{m'}\right)&=& \frac{\vec{v}_{m'}^{\,T}\left(\vec{\phi}^{\prime}+\frac{1}{\sigma_{m}^{*}}\vec{\phi}^{*\prime}\right)}{\vec{v}_{m'}^{\,T}\left(\vec{\phi}+\frac{1}{\sigma_{m}^{*}}\vec{\phi}^{*}\right)}=\nonumber\\
&=&\label{lambda-rew-above}\frac{\vec{v}_{m}^{\,\dagger}P\left(\vec{\phi}^{\prime*}+\sigma_{m}^{*}\vec{\phi}^{\prime}\right)}{\vec{v}_{m}^{\,\dagger}P\left(\vec{\phi}^{*}+\sigma_{m}^{*}\vec{\phi}\right)}
 = \mathcal{P}\lambda^{*}\!\left(\sigma_{m}\right).
\end{eqnarray}
From Eqs.(\ref{lambda-rew-above}) and (\ref{lambda-rew-below}) we may thus conclude, that for 2D $\mathcal{PT}$--systems the situation 
is very similar to the 1D case: Below the $\mathcal{PT}$-threshold of a pair of $S$-eigenvalues, the corresponding eigenstates feature different, $\mathcal{PT}$-symmetric BC, above threshold the BC are non-$\mathcal{PT}$-symmetric, but pairwise connected.

\end{appendix}

\bibliographystyle{apsrev-no-url}

\end{document}